\documentclass[aps,prl,twocolumn,showpacs,superscriptaddress]{revtex4}

\usepackage{graphicx}
\usepackage{dcolumn}
\newcommand\degrees[1]{\ensuremath{#1^\circ}}

\begin{document}

\title{Ferrodistortive instability at the (001) surface of half-metallic 
manganites}

\author{J. M. Pruneda}
\affiliation{Department of Physics, University of California,
Berkeley, California 94720} 
\affiliation{Institut de Ci\`encia de Materials de Barcelona, 
(ICMAB-CSIC), Campus de la UAB, 08193 Bellaterra, Spain}
\author{V. Ferrari}
\affiliation{Cavendish Laboratory, University of Cambridge, 
Madingley Road, Cambridge CB3 0HE, UK}
\affiliation{Departamento de F\'{\i}sica, Comisi\'on Nacional de 
Energ\'{\i}a At\'omica, Gral. Paz 1499, 1650 San Mart\'{\i}n, 
Buenos Aires, Argentina.}
\author{P. B. Littlewood}
\affiliation{Cavendish Laboratory, University of Cambridge, 
Madingley Road, Cambridge CB3 0HE, UK}
\author{N. A. Spaldin}
\affiliation{Materials Department, University of
California, Santa Barbara, California 93106-5050}
\author{Emilio Artacho}
\affiliation{Department of Earth Sciences, University of Cambridge, 
Downing Street, Cambridge CB2 3EQ, UK}

\date{\today}

\begin{abstract}
We present the structure of the fully relaxed (001) surface of the 
half-metallic manganite La$_{0.7}$Sr$_{0.3}$MnO$_3$, calculated 
using density functional theory within the generalized gradient 
approximation (GGA).
Two relevant ferroelastic order parameters are identified and 
characterized: The tilting of the oxygen octahedra, which is present in 
the bulk phase, oscillates and decreases towards the surface, and 
an additional ferrodistortive Mn off-centering, triggered by the 
surface, decays monotonically into the bulk.
The narrow $d$-like energy band that is characteristic of unrelaxed
manganite surfaces is shifted down in energy by these structural 
distortions, retaining its uppermost layer localization.
The magnitude of the zero-temperature magnetization is unchanged from
its bulk value, 
but the effective spin-spin interactions are reduced at the surface.
\end{abstract}

\maketitle

A good understanding of the surface physical properties in colossal 
magneto-resistant hole-doped manganese oxides is highly desirable for 
future applications of these promising materials in magnetoresistive 
devices or spintronics.~\cite{Dorr} 
The metallic phase of  La$_{1-x}$A$_x$MnO$_3$ ($A=$Sr, Ca), obtained for 
$x\sim 0.3$ is of particular interest because of its high spin polarization 
at the Fermi level.
  There is some controversy, however, on whether this optimally doped phase 
is a half-metal, with a gap for one spin direction, and thus with fully
spin-polarized carriers.
  Early spin-resolved photoemission experiments observed such ideal 
polarization \cite{Park}, although electron tunneling 
experiments show traces of partial occupation of minority spin
states \cite{Nadgorny}.
Since these techniques probe the bulk structure through the surface or 
interface, (as do most techniques for bulk studies with variable degrees
of surface sensitivity), the origin of these discrepancies could lie not 
only in the difficulty in preparing smooth surfaces at an atomic level, 
but also in the fact that the properties of the surface may markedly 
deviate from those of the bulk. The primary goal of this work is to 
determine the structural changes induced by surface termination at the 
(100) surface of La$_{1-x}$Sr$_x$MnO$_3$ (LSMO). In addition, we discuss
the influence of these structural changes on the electronic properties.

The theoretical description of the half-metallic manganites is challenging
given the strong correlations of the transition metal $d$ electrons,
as well as the complex chemistry and large range of possible structural 
distortions from the high symmetry perovskite structure. For example,
{\it ab initio} studies based on the local spin density approximation 
(LSDA) give poor agreement with the experimental lattice constant; as a 
result many first principles calculations to date have been performed for 
cubic structures with the experimental pseudocubic lattice constant 
and without structural relaxations. 
However, the strong electron-lattice interactions are known to play an 
important role in the electronic properties, and a full description of 
the lattice distortions, particularly at surfaces or interfaces where local 
distortions might be more pronounced, is desirable.
Furthermore, LSDA fails to obtain a half-metallic state for $x=0.3$
\cite{PickettSingh}, and it has been argued that a more accurate 
treatment of electron correlation, such as LSDA$+U$ or self-interaction 
corrected (SIC) is required, particularly for the
charge-ordered or insulating phases \cite{Satpathy,Solovyev}.

Here we present {\it ab initio} results for a fully relaxed LSMO (001) 
surface obtained using density-functional theory within the generalized
gradient approximation (GGA). Our choice of exchange-correlation functional 
was motivated by our recent GGA calculations for bulk
La$_{0.7}$Sr$_{0.3}$MnO$_3$ (LSMO),~\cite{Ferrari} in which we found good
agreement between the theoretically optimized pseudocubic and experimental
lattice constants, and qualitatively correct half-metallic electronic 
behavior. Since our focus is on structural, rather than on electronic 
properties, and since we are interested in the metallic phase of LSMO, 
which reduces the errors related to the non-locality of the 
exchange-correlation hole, we elect not to include higher order 
descriptions of correlation effects such as in the GGA$+U$ method.  Indeed, 
it was recently shown \cite{GGA+U} that the use of GGA$+U$ does not 
significantly change the structural properties.  
However the limitations of today's GGAs for treating highly-correlated 
systems should be kept in mind, in particular when we discuss changes 
in the surface electronic spectral properties based on Kohn-Sham 
eigenvalues \cite{limitations}. 

  Our calculations were performed using the {\sc Siesta} implementation
of density functional theory 
\cite{SIESTAidea, SIESTAmethod}.
  Core electrons were replaced by norm-conserving pseudopotentials 
\cite{Troullier-Martins}, and finite-support numerical atomic orbitals 
were used to describe the valence electrons, at the double-$\zeta$
polarized level \cite{Anglada}. 
  Semicore states for Ti ($3s^2$ and $3p^6$), Sr ($4s^2$ and $4p^6$), 
and La ($5s^2$ and $5p^6$) were included in the valence.
  The spin-polarized gradient-based PBE exchange-correlation functional was 
used \cite{PBE}, with a grid cutoff of 320 Ry for integrations in real space.
  Atomic positions were relaxed until the forces were smaller than 25 meV/\AA. 

  To validate our method in obtaining relaxed atomic structures, we first 
studied the properties of bulk LSMO.
  We built a cell that is a $\sqrt{2}\times\sqrt{2}\times 2$ repetition of 
the ideal perovskite cubic cell, and performed a variable cell relaxation 
of the crystal structure.   
  Our calculated lattice parameters are within 2 \% of the experimentally 
reported values, and our Mn-O bond lengths (1.96 \AA) are even closer to
experiment (1.95 \AA) \cite{Radaelli}.
The tilt angle (\degrees{6.6}) that we obtain
for the optimized structure is in excellent agreement with the 
experimental value (\degrees{6.8} \cite{Radaelli}).
Since these distortions are believed to be the source of the strong 
electron-phonon coupling in manganites~\cite{Millis} (they lift the 
degeneracy of $e_g$ levels, as recently shown in photoemission 
experiments \cite{Falub05}, and reduce the electronic energy 
for single occupancy of this state) their accurate representation in
our calculations is essential.

  To study the surface, we consider a supercell structure with 4$\times$2 
repetitions of the original cubic cell in the surface $ab$-plane, eleven atomic 
layers (5.5 unit cells) along the direction perpendicular to the surface, 
and MnO$_2$ termination at both sides of the slab \cite{MnOtermination}.
  We begin with a pseudocubic structure and allow for atomic relaxations, 
while keeping the in-plane lattice parameters fixed to the relaxed bulk values.
  The supercell is large enough to accommodate the octahedral tiltings,
and commensurate with the orthorhombic phase in the $ab$ plane.
  The surface breaks the bulk rhombohedral symmetry, most importantly the
three-fold axes of the $R\bar3c$ phase, therefore falling back onto the 
$Pnma$ tilt system.
  Doping is then introduced explicitly by substituting La atoms by Sr, with 
the resulting structure having a quasi-random distribution of 13 Sr atoms
and 27 La atoms.
  The system is thick enough to ensure that the electronic structure at the 
center of the slab is similar to that of the bulk.  

\begin{figure}[t!]
\includegraphics[width=0.42\textwidth]{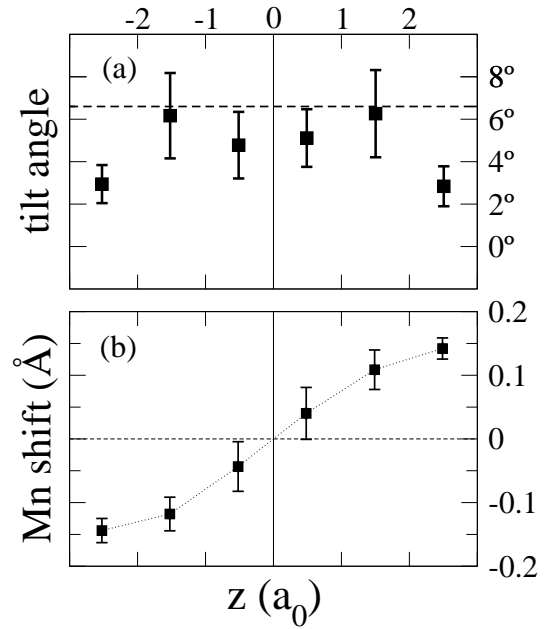}%
\caption[]{\label{fstructure} (a) Octahedral tilt angle versus position 
across the slab, $z$, in units of the cubic lattice constant, $a_0$.
  The dashed line shows the value for the theoretically relaxed bulk geometry.
  (b) Off-centering (in \AA) of the Mn atoms from the plane 
defined by the neighbouring O atoms in the MnO$_2$ layer.
  The $z$ origin is defined at the center of the slab.
  Error bars indicate the differences within the same atomic layer.}
\end{figure}

We find that relaxation of the atomic positions gives rise to a buckling in the 
surface layer, with the Mn atoms shifted out of the plane.
  There is also a change in the octahedral tiltings and of the Mn-O bonding 
distances.
  In Fig.~\ref{fstructure}(a) we plot the tilting angle of the MnO$_6$ 
octahedra as a function of the position of the Mn layer across the slab.
  It can be seen that in the central layers, the tilting approaches the value 
obtained for the relaxed bulk structure (dashed line) while it reduces 
towards the surface.
In addition, we find that
  the buckling of the Mn atoms with respect to oxygens in the surface 
layer propagates to the subsurface giving rise to an off-centering 
similar to that of ferroelectric perovskites.
  The magnitude of this off-centering, shown in Fig.~\ref{fstructure}(b), 
decays monotonically towards the center of the slab, where it is zero 
due to the cancelling effect of both surfaces.  The distortion results in 
important changes in the Mn-O bonding distances along the $z$ direction, 
whereas the in-plane bonds remain essentially unchanged.  

  Although the Mn-O buckling at the surface layer is a result of
local surface chemistry (see the surface-state shift below), the 
off-centering deeper into the bulk is not.
  The situation is better described as a polar, ferroelastic instability, 
which is triggered by the surface buckling. Since this instability is not 
dominant in the bulk phase, it therefore decays from the surface into the bulk.
Such surface-triggered decaying ferroelastic distortions are described 
in the literature \cite{Ekhard}, and can be very slowly decaying
\cite{Hayward}.
Our calculations, however, allow us to ascertain only that the decay length 
$\lambda_o$ is not smaller than $\sim$10\AA, since the antisymmetric 
character of the order parameter [Fig.~\ref{fstructure}(b)] forces it 
to go through zero at the center of the slab.

  The length-scale of the decay towards the surface of the octahedral tilt 
order parameter, $\lambda_t$, should be captured in our calculations, 
since the symmetry of the slab does not constrain it in this case.
We find that it is of the order of a few atomic layers.
  However, we can not rule out the possibility that both order parameters 
are coupled at the surface, whereby $\lambda_t$ would be in fact determined by the constraint on
$\lambda_o$. Indeed, couplings and/or competitions between polar ferroelectric
modes and tiltings and rotations of oxygen octahedra are well known to occur
in related perovskites \cite{Woodward_1:1997,Woodward_2:1997,Bilc/Singh:2006} 
  Interestingly, however, the decay of the untilting seems oscillatory and 
commensurate with the crystal; both monotonic and oscillatory behaviors can 
be expected in surface ferroelastic decays \cite{Ekhard}.

\begin{figure}[t!]
\includegraphics[width=0.42\textwidth]{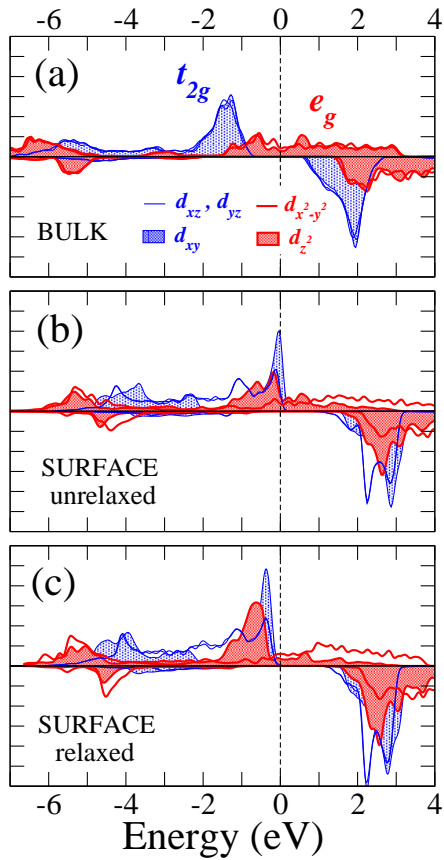}
\caption[]{\label{pdos} (Color online): 
Partial density of states for up/down spins 
(top/bottom) for the Mn $3d$ orbitals in the center of the slab (a), 
and the surface layer before (b) and after (c) the structural relaxation.
  The surface state is mainly composed of $d_{z^2}$ and $d_{xy}$ 
orbitals, which are shown as shaded regions.}
\end{figure}

Although not the main focus of this work, next we briefly compare the 
bulk and surface electronic structures within the GGA.
Fig.~\ref{pdos} shows our calculated Mn $3d$ partial densities of states 
(PDOS) in the center of the slab (a), and for unrelaxed (b) and relaxed 
(c) surface configurations.
At the center of the slab, the system is half-metallic, with $e_g$ character 
for the states at $\varepsilon_F$, and localized $t_{2g}^\uparrow$ states around
1.5~eV below, in fair agreement with previous calculations \cite{PickettSingh}
and photoemission experiments \cite{Chikamatsu06}.  
Importantly, this half-metallicity persists at the surface, and since the
majority spin metallicity results from a broad Mn $3d$ - O $2p$ hybridized
band (the Mn $d_{z^2}$ and $d_{x^2-y^2}$ states, 
as well as the $p_z$ of the oxygen underneath the Mn, dominate the Fermi
level at the surface)  
it is likely to be accurately represented in our GGA methodology.
  For the unrelaxed (cubic) structure, the surface termination induces a 
localized surface state, of mainly $d_{z^2}$ and $d_{xy}$ character. 
This state had been previously observed in LSDA and SIC-LSDA calculations 
with unrelaxed cubic surface geometries, and originates after the splitting 
of the states $e_g \rightarrow d_{z^2},d_{x^2-y^2}$ and 
 $t_{2g}\rightarrow d_{xy},(d_{yz},d_{xz})$ related to the surface-induced 
symmetry breaking, from cubic $O_h$ to $C_{4v}$~\cite{Filippetti,Zenia}.
  The surface state shifts to lower energies away from $\varepsilon_F$ 
when the geometry relaxes. This is the driving force for the surface buckling.

  Concerning magnetic properties, we observe that the zero kelvin magnetic moment 
remains almost unaltered up to the very surface, with a value
of 3.55$\mu_B$ at the relaxed surface compared with 3.53$\mu_B$ in the bulk.
  The ground state remains ferromagnetic.  In order to address the finite 
temperature magnetism, we explored different magnetic configurations and 
found that the energetic 
cost of reversing one magnetic moment at the surface is reduced to 30\% of
the bulk value, indicative of lower effective magnetic interactions,
and in line with experimental results~\cite{Park}.  
Forces on the atoms of up to 1.54 eV/\AA\ (1.01 eV/\AA\ ) have been found 
upon single spin reversal in the surface (in the bulk), 
indicating (i) that the surface ferromagnetic 
coupling would weaken even more as compared to the bulk, and 
(ii) a possible important magnon-phonon coupling.

  Two competing considerations arise when relating the magnetic
results to the structural ones. The double-exchange (DEX) mechanism, 
in which localized 3$d$ $t_{2g}$ electron 
spins interact ferromagnetically through the $e_g$ conduction-electron 
hopping between nearest neighbours, is frequently used to describe the 
electronic properties of manganites.  
  On one hand, the electron-hopping energy among Mn $e_g$ states in this
DEX model is expected to decrease with the bending of the 
Mn-O-Mn bonds, in agreement with the observed reduction of the 
ferromagnetic Curie temperature $T_M$ in doped manganites~\cite{Fontcuberta}.
  The octahedral untilting towards the surface discussed before would 
thus suggest a surface enhancement of hopping and $T_M$, at odds with
the mentioned observations.
On the other hand, however, the hopping energy should be much more 
affected by the substantial off-centering, thus altering the inter-layer 
coupling.

  The surface-triggered ferrodistortive instability discussed above is very 
similar in kind and magnitude to that of ferroelectrics like BaTiO$_3$.
  It is thus tempting to consider multiferroic possibilities for this system. 
Considering the calculated values for the atomic displacements, 
and taking $\sim+3e$ for the nominal charge of Mn, we would obtain 
a polarization of $\sim$10$\mu$C/cm$^2$ at the surface layer, 
comparable to the values obtained in similar Mn-based multiferroics.
  It must be remembered, however, that it is a metallic system, and thus
any polarization will be screened by the itinerant carriers. 
  In addition to the ferroelastic characteristics discussed above, 
the surface buckling could display surface piezoelectric signatures,
considering the Thomas-Fermi screening length of this (poor) metal,
affecting the work function and its strain dependence.  
  This could be a relevant consideration for understanding
the orientation of dipolar molecules deposited on the surface.
  Spin valve devices based in LSMO have already been demonstrated with 
Alq3 (8-hydroxy-quinoline aluminum) \cite{organicvalves}, and a 
substantial shift in its molecular energy levels has been 
measured~\cite{Riminucci07}.
  The LSMO surface dipole would favor the orientation of the Alq3 molecular 
dipole towards the vacuum side, and the resulting electrostatic potential 
would lower the molecular levels as observed~\cite{Riminucci07}.

  It should be remembered that the presented results here are for the 
defect-free MnO$_2$-terminated LSMO surface.  
They may be very different for (La,Sr)O termination~\cite{Zenia}, 
and, most importantly, in the presence of defects like oxygen vacancies.
  Recent photoemission experiments have suggested the presence of strongly 
localized Mn$^{+2}$ states at the surface of LSMO, related to
oxygen vacancies \cite{deJong06}. 
  DFT results on the effect of these defects in this LSMO surface will be
discussed elsewhere \cite{Rurali}.

  In conclusion, our first-principles structural relaxation of the LSMO
(001) surface has revealed a surprising surface-triggered ferroelastic
instability of an off-centering ferrodistortive kind.
  The distortion is driven by the surface electronic structure, and 
causes a weakening of the tendency to magnetic order close to the surface.
  We hope that our observations will stimulate further work towards 
the characterization of the discussed instabilities, as well as possible
exploitation of the newly observed surface polarization and its coupling
to the magnetic behavior. 

We acknowledge discussions with E. K. H. Salje, L. E. Hueso, M. A. Carpenter,
L. Brey and N. D. Mathur. The calculations were done at the HPCF of Cambridge
University. This work was funded by the UK EPSRC, CONICET-Argentina (VF), 
NERC, BNFL, the UE Marie Curie Fellowship (JMP), and the National Science 
Foundation's Division of Materials Research, grant number DMR-0605852 (NAS). 
NAS thanks the Miller Institute at UC Berkeley for their support during this work.


\end{document}